\documentclass{PoS}
\usepackage{slashed}
\usepackage{amssymb}
\usepackage{amsmath}
\usepackage{amsfonts}

\newcommand{\be}{\begin{equation}}
\newcommand{\ee}{\end{equation}}
\newcommand{\bea}{\begin{eqnarray}}
\newcommand{\eea}{\end{eqnarray}}

\title{Novel Scenarios for Majorana Neutrino Mass Generation and Leptogenesis from Kalb-Ramond Torsion.}

\ShortTitle{Kalb-Ramond Torsion and Neutrinos}

\author{\speaker{Nikolaos E. Mavromatos}\thanks{Currently also at: Theory Division, Department of Physics, CERN, CH 1211 Geneva 23, Switzerland.
This work is supported in part by the London Centre for Terauniverse Studies (LCTS), using funding from the European Research Council via the Advanced Investigator Grant 267352, and by STFC (UK) under the research grant ST/L000326/1.}\\
        Theoretical Particle Physics and Cosmology Group, Department of Physics, King's College London, Strand, London WC2R 2LS, UK\\
        E-mail: \email{Nikolaos.Mavromatos@kcl.ac.uk}}

\abstract{The Kalb-Ramond (KR) antisymmetric tensor  field arises naturally in the gravitational multiplet of string theory. Nevertheless, the respective low-energy field theory action, in which, for reasons of gauge invariance, the only dependence on the KR field is through its field strength,  constitutes an interesting model \emph{per se}. In this context, the KR field strength also acts as a totally antisymmetric torsion field, while in four space-time dimensions is \emph{dual} to an (KR) axion-like pseudoscalar field. In this context, we review here first the r\^ole of quantum fluctuations of the KR axion on the generation of Majorana mass for neutrinos, via a mixing with ordinary axions that may exist in the theory as providers of dark matter candidates. Then we proceed to discuss the r\^ole of constant in time (thus Lorentz violating) KR torsion backgrounds, that may exist in the early Universe but are completely negligible today, on inducing Leptogenesis by means of \emph{tree-level} CP violating decays of Right Handed Massive Majorana neutrinos in the presence of such H-torsion backgrounds. Some speculations regarding microscopic D-brane world models, where such scenarios may be realised, are also given.}

\FullConference{18th International Conference From the Planck Scale to the Electroweak Scale \\
		25-29 May 2015\\
		Ioannina, Greece }

\begin{document}

\section{Introduction and Summary}

The  discovery~\cite{higgs} of  the Higgs  boson at  the CERN
Large Hadron Collider (LHC) in 2012 constitutes an important milestone for the
Ultra-Violet (UV) completion of the Standard Model (SM), verifying that a  Higgs-like mechanism  can explain the generation  of most of
the particle masses in the SM. Necvertheless, the origin of the small neutrino masses
still remains an open issue.  One such mechanism providing a rather  natural
explanation of the observed smallness of
the  light neutrino  masses  is the so-called \emph{see-saw}  mechanism~\cite{seesaw},   which  necessitates  the  Majorana
nature of the light (active)  neutrinos and postulates the presence of
heavy right-handed  Majorana partners of mass  $M_R$. The right-handed
Majorana mass $M_R$  is usually considered to be  much larger than the
lepton or  quark masses.  The  origin of $M_R$  has been the  topic of
several extensions of  the SM in the literature,  within the framework
of quantum field  theory~\cite{seesaw,Schechter:1980gr}    
and string theory~\cite{Blumenhagen:2006xt}. However, up to now, there is no experimental evidence for right-handed neutrinos or, in fact, for any extension of the SM, although some optimism of discovering supersymmetry in the current round of LHC (operating ultimately at 14 TeV energies)
exists among particle physicists~\cite{mitsoususy}. 

Until therefore such extensions of the SM are discovered, it is legitimate to search for alternative mechanisms for neutrino mass generation, 
that keep the spectrum of SM intact, except perhaps for the existence of right handed neutrinos that are allowed. 
Such minimal, non supersymmetric extensions of the Standard Model with three in fact right-handed Majorana neutrinos complementing the three active 
left-handed neutrinos  (termed $\nu$MSM), have been proposed~\cite{nuMSM}, in a way consistent with current cosmology. Such models are characterised by relatively light right-handed neutrinos, two of which are almost degenerate, with masses of order GeV, and a much lighter one, almost decoupled, with masses in the keV range, which may play the role of warm dark matter. The right-handed neutrinos serve the purpose of generating, , through seesaw type mechanisms, the active neutrino mass spectrum, consistent with observed flavour oscillations. However, there are no suggestions for microscopic mechanisms for the generation of the right-handed neutrino mass spectrum in such scenarios. 

Motivated by these facts we first review in this talk, in section \ref{sec:torsion}, an alternative proposal for dynamical generation of Majorana mass for neutrinos 
propagating in space-time geometries with quantum-fluctuating torsion~\cite{mptorsion}.  Microscopically this torsion may be provided by the spin-one antisymmetric tensor (Kalb-Ramond (KR)) field~\cite{Kalb} of the gravitational multiplet of a string~\cite{tseytlin,kaloper}, but the low energy theory of the KR torsion is an interesting field theory model that can be studied in its own right.  In this scenario, the totally antisymmetric part of the torsion couples, via the gravitational covariant derivative, to all fermions in a way that the resulting interaction resembles that of the Lorentz and CPT-Violating pseudovector background with the axial fermion current in the Standard Model Extension (SME) of Kostelecky and collaborators~\cite{kostel}. The generation of (right-handed, sterile) neutrino masses in that case proceeds, as we shall review below, via chiral anomalous three-loop graphs of neutrinos interacting with the totally antisymmetric torsion quantum field. In four space-time dimensions, the latter is represented as an axion field~\cite{tseytlin,kaloper}, whose mixing with ordinary axion fields, that in turn interact with the Majorana right-handed neutrinos via chirality changing Yukawa couplings, is held responsible for the right-handed Majorana neutrino mass generation through the aforementioned anomalous graphs. 
In the second part of the talk, in section \ref{sec:lepto}, we discuss non-trivial backgrounds of the KR torsion field, constant in cosmic time, and thus Lorentz-violating, whose coupling to the fermion axial current induces extra sources of CP violation, which  in the case of right-handed Majorana neutrinos assumed abundant in the early Universe, may lead to Leptogenesis~\cite{sarkarlepto1,sarkarlepto}. 
Finally, in the last part of the talk, section \ref{sec:brane}, instead of conclusions, I present some speculations as to how the above scenarios of Leptogenesis, which require relatively large backgrounds of the KR torsion in early epochs of the Universe, but negligible today, due to phenomenological reasons, may be realised in some microscopic string/brane models of the Universe, in which our four space-time dimensional brane world propagates in a bulk punctured by populations of point-like D-brane defects, interacting with right-handed neutrinos~\cite{mavromatosDfoam}. Some technical aspects of our approach are outlined in an Appendix.

\section{Neutrino Mass Generation due to Quantum (Kalb-Ramond) Torsion\label{sec:torsion}}

Let us for concreteness 
consider  Dirac fermions in  a torsionful  space-time~\cite{torsion}.  The extension to the Majorana case is straightforward. 
The relevant 
action reads~\footnote{The second term in the right-hand-side of (\ref{dirac})  is usually not written in flat space, as its contribution is equal to the first term plus a surface integral. However, the situation is different when spacetime is not flat. In fact the second term is needed in order preserve unitarity, allowing for the cancellation of an anti-hermitean term involving the trace of the spin-connection coefficients $\overline \omega^{a}_{\;ac}$.}:
\begin{equation}\label{dirac}
S_\psi = \frac{i}{2} \int d^4 x \sqrt{-g} \Big( \overline{\psi}
\gamma^\mu \overline{\mathcal{D}}_\mu \psi  
- (\overline{\mathcal{D}}_\mu \overline{\psi} ) \gamma^\mu \psi \Big)
\end{equation}
where $\overline{\mathcal{D}}_\mu = \overline{\nabla}_\mu  + \dots $,
is the covariant derivative (including gravitational and gauge-field connection parts (the latter represented by $\dots$), in case the fermions are charged). The
overline above  the covariant derivative, i.e.~$\overline{\nabla}_\mu$,
denotes  the presence  of  torsion, 
\be
\overline \nabla_{a}\psi=e_{a}^{\mu}\left(\partial_{\mu}+\frac{i}{2}\overline \omega_{b\mu c}\Sigma^{bc}\right)\psi.
\ee
In the formula above $\Sigma^{ab}=\frac{i}{4}\left[\gamma^{a},\gamma^{b}\right]$ is the generator of the Lorentz group representation on four-spinors, while $\overline\omega_{a\mu b}$ is the torsionful spin connection
\be\label{omegabar}
\overline \omega^{a\; b}_{\, \mu}=e^{a}_{\nu}\overline\nabla_{\mu}e^{b\nu}=e^{a}_{\nu}\left(\partial_{\mu}e^{b\nu}+\overline{\Gamma}^{\nu}_{\mu\lambda}e^{b\lambda}\right).
\ee
with $\overline \Gamma^\mu_{\,\,\nu\rho} \ne \overline \Gamma^\mu_{\,\,\rho\nu} $ the generalised Christoffel symbols in the presence of torsion. 

The torsionful spin connection can be decomposed as: $\overline{\omega}_{a  b \mu} = \omega_{a b
  \mu} + K_{a  b \mu} $, where $K_{ab \mu}$  is the contorsion tensor.
The  latter  is  related  to  the  torsion  two-form  $\textbf{T}^a  =
\textbf{d   e}^a   +    \overline{\omega}^a   \wedge   \textbf{e}^b   $
via~\cite{torsion,kaloper}:     $K_{abc}    =     \frac{1}{2}    \Big(
\textrm{T}_{cab}  - \textrm{T}_{abc}  -  \textrm{T}_{bcd} \Big)$.  The
presence  of torsion  in the  covariant derivative  in  the 
action (\ref{dirac}) leads, apart from the standard terms in manifolds
without  torsion, to an  additional term  involving the  axial current
\be\label{axial}
J^\mu_5 \equiv \overline{\psi} \gamma^\mu \gamma^5 \psi~.
\ee
The relevant part of the action reads:
\begin{equation}\label{torsionpsi}
S_\psi \ni  - \frac{3}{4} \int d^4 \sqrt{-g} \, S_\mu \overline{\psi}
\gamma^\mu \gamma^5 \psi  = - \frac{3}{4} \int S \wedge {}^\star\! J^5  
\end{equation}
where $\textbf{S} = {}^\star\! \textbf{T}$  is the dual of \textbf{T}: $S_d
=   \frac{1}{3!}     \epsilon^{abc}_{\quad   d}   T_{abc}$.     

We  next remark that  the torsion  tensor can  be decomposed  into its
irreducible parts~\cite{torsion},  of which $S_d$  is the pseudoscalar
axial vector:
$ T_{\mu\nu\rho} = \frac{1}{3} \big(T_\nu
g_{\mu\rho} - T_\rho g_{\mu\nu} \big) - \frac{1}{3!}
\epsilon_{\mu\nu\rho\sigma} \, S^\sigma + q_{\mu\nu\rho}$, 
with
$\epsilon_{\mu\nu\rho\sigma} q^{\nu\rho\sigma} = q^\nu_{\,\rho\nu} =
0$.
This implies that the contorsion tensor undergoes the following decomposition:
\begin{equation}\label{hatted}
K_{abc} = \frac{1}{2} \epsilon_{abcd} S^d + {\widehat K}_{abc} 
\end{equation}
where $\widehat  K$ includes the  trace vector $T_\mu$ and  the tensor
$q_{\mu\nu\rho}$ parts of the torsion tensor.
 
The gravitational part of the action can then be written as:
$
S_G =\frac{1}{2\kappa^2} \, \int d^4 x \sqrt{-g} \Big(R +
\widehat{\Delta} \Big) + \frac{3}{4\kappa^2} \int \textbf{S} \wedge
{}^\star\! \textbf{S},$
where  $\widehat \Delta  = {\widehat  K}^\lambda_{\ \: \mu\nu} {\widehat
  K}^{\nu\mu}_{\quad \lambda}  - {\widehat K}^{\mu\nu}_{\quad  \nu} \,
{\widehat K}^{\quad  \lambda}_{\mu\lambda}$, with the  hatted notation
defined in (\ref{hatted}).

In a  quantum gravity setting,  where one integrates over  all fields,
the torsion terms  appear as non propagating fields  and thus they can
be integrated out exactly. The authors of \cite{kaloper} have observed
though   that  the   classical  equations   of  motion   identify  the
axial-pseudovector torsion field $S_\mu$ with the axial current, since
the torsion equation yields
\begin{equation}\label{torsionec}
K_{\mu a b} = - \frac{1}{4} e^c_\mu \epsilon_{a b c d} \overline{\psi}
\gamma_5 {\tilde \gamma}^d \psi\ .
\end{equation}
From this  it follows $\textbf{d}\,{}^\star\!\textbf{S}  = 0$, leading
to a  conserved ``torsion charge'' $Q =  \int {}^\star\!  \textbf{S}$.
To  maintain  this conservation  in  quantum  theory, they  postulated
$\textbf{d}\,{}^\star\!\textbf{S} = 0$ at the quantum level, which can
be  achieved  by  the  addition  of  judicious  counter  terms.   This
constraint, in a path-integral formulation of quantum gravity, is then
implemented  via a delta  function constraint,  $\delta (d\,{}^\star\!
\mathbf{S})$, and the latter via the well-known trick of introducing a
Lagrange multiplier  field $\Phi (x)  \equiv (3/\kappa^2)^{1/2} b(x)$.
Hence, the relevant torsion  part of the quantum-gravity path integral
would include a factor {\small
\begin{eqnarray}
 \label{qtorsion}
&&\hspace{-5mm} \mathcal{Z} \propto \int D \textbf{S} \, D b   \, \exp \Big[ i \int
    \frac{3}{4\kappa^2} \textbf{S} \wedge {}^\star\! \textbf{S} -
      \frac{3}{4} \textbf{S} \wedge {}^\star\! \textbf{J}^5  +
      \Big(\frac{3}{2\kappa^2}\Big)^{1/2} \, b \, d {}^\star\! \textbf{S}
      \Big]\nonumber \\  
&&\hspace{-5mm}=\!  \int D b  \, \exp\Big[ -i \int \frac{1}{2}
      \textbf{d} b\wedge {}^\star\! \textbf{d} b + \frac{1}{f_b}\textbf{d}b 
\wedge {}^\star\! \textbf{J}^5 + \frac{1}{2f_b^2}
    \textbf{J}^5\wedge\, ^\star \textbf{J}^5 \Big]\; ,\nonumber\\
\end{eqnarray}
\hspace{-1.5mm}}
where 
$f_b = (3\kappa^2/8)^{-1/2} = \frac{M_P}{\sqrt{3\pi}}$ 
and  the  non-propagating   $\textbf{S}$  field  has  been  integrated
out. The reader  should notice that, as a  result of this integration,
the   corresponding   \emph{effective}   field   theory   contains   a
\emph{non-renormalizable} repulsive four-fermion axial current-current
interaction, characteristic of any torsionful theory~\cite{torsion}.

The torsion term, being geometrical, due to gravity, couples universally to all fermion species, not only neutrinos.
Thus, in the context of the SM of particle physics,  the axial current (\ref{axial}) is expressed as a sum over fermion species
\be\label{axial2}
J^\mu_5 \equiv \sum_{i={\rm fermion~species}} \, \overline{\psi}_i  \gamma^\mu \gamma^5 \psi_i ~.
\ee
In theories with chiral anomalies, like the quantum electrodynamics part of SM, 
the  axial current  is not
conserved at the  quantum level, due to anomalies,  but its divergence
is obtained by the one-loop result~\cite{anomalies}:
\begin{eqnarray}
   \label{anom}
\nabla_\mu J^{5\mu} \!&=&\! \frac{e^2}{8\pi^2} {F}^{\mu\nu}
  \widetilde{F}_{\mu\nu}  
- \frac{1}{192\pi^2} {R}^{\mu\nu\rho\sigma} \widetilde
{R}_{\mu\nu\rho\sigma} \nonumber\\ 
&\equiv& G(\textbf{A}, \omega)\; .
\end{eqnarray}
We  may then partially integrate  the second  term in  the exponent  on the
right-hand-side  of (\ref{qtorsion})  and take  into account (\ref{anom}).
The reader should observe that in (\ref{anom}) the torsion-free spin connection has been
used.  This can be achieved by the addition of proper counter terms in
the  action~\cite{kaloper}, which  can  convert the  anomaly from  the
initial    $G(\textbf{A},   \overline   \omega)$    to   $G(\textbf{A},
\omega)$. Using  (\ref{anom}) in (\ref{qtorsion}) one  can then obtain
for the effective torsion action in theories with chiral anomalies, such as the QED part of the SM:
\begin{equation}\label{brr}
\int D b\ \exp\Big[ - i \int \frac{1}{2}
    \textbf{d} b\wedge {}^\star\! \textbf{d} b  - \frac{1}{f_b} b
    G(\textbf{A}, \omega)  
+ \frac{1}{2f_b^2} \textbf{J}^5 \wedge \, ^\star \textbf{J}^5 \Big]\; .
\end{equation}

A concrete example of torsion  is provided by string-inspired theories, where the totally antisymmetric component $S_\mu$ of the torsion 
is identified with the field strength of the spin-one antisymmetric tensor (Kalb-Ramond (KR)~\cite{Kalb}) field
$H_{\mu\nu\rho} = \partial_{[\mu} B_{\nu\rho]}$, 
where  the  symbol  $[\dots   ]$  denotes  antisymmetrization  of  the
appropriate indices.
The  string theory effective action depends only on $H_{\mu\nu\rho}$ as a consequence of the ``gauge symmetry''
$B_{\mu\nu} \rightarrow B_{\mu\nu} + \partial_{[\mu }\Theta_{\nu]} $ that characterises all string theories. 
Indeed, in the Einstein frame, to first order in the string Regge $\alpha^\prime$, the bosonic part of the low-energy effective action (in four large target-space-time dimensions)  is given by \cite{tseytlin}
\be\label{low energy effective action}
S=\frac{1}{2\kappa^2}\int \mbox{d}^4x\;\sqrt{-g}\left(R-2\partial^{\mu}\phi\partial_{\mu}\phi-e^{-4\phi}H_{\lambda\mu\nu}H^{\lambda\mu\nu}- V(\phi)\right),
\ee
where $\frac{1}{\kappa^2} \equiv \frac{M_{s}^2\Omega^{c}}{8\pi} = \frac{1}{8\pi {\rm G}}$, with ${\rm G}$ the four-dimensional (gravitational) constant, $M_{s}^2=1/\alpha^\prime$ is the string mass scale, $\Omega^{c}$ the (dimensionless) compactification volume in units of the Regge slope $\alpha^\prime$ of the string and $V(\phi)$ is a dilaton potential.

It can be shown~\cite{tseytlin} that the
terms of the effective action 
up to and including quadratic order in the Regge slope parameter $\alpha^\prime$, of relevance to the low-energy (field-theory) limit of string theory, 
which involve the H-field strength,
can  be assembled in such a way
that  only    torsionful     Christoffel    symbols, $\overline{\Gamma}^\mu_{\nu\rho}$ 
appear:
\be\label{connection}
\overline{\Gamma}^{\lambda}_{\;\mu\nu}=\Gamma^{\lambda}_{\;\mu\nu}+e^{-2\phi}H^{\lambda}_{\;\mu\nu} \ne \overline{\Gamma}^{\lambda}_{\;\nu\mu}.
\ee
where $\Gamma^\mu_{\nu\rho}  = \Gamma^\mu_{\rho\nu}$ is  the ordinary,
torsion-free, symmetric connection,  and $\kappa$ is the gravitational
constant. In four space-time dimensions, the dual of the H-field is indeed the derivative of an axion-like field, 
\be\label{bfield}
 \partial^\mu b = -\frac{1}{4} \, \varepsilon^{abcd}\, e^{-2\phi}\, H_{abc}~.
\ee
where $b(x)$ is a pseudoscalar field, entirely analogous to the field $b$ above, hence the use of the same letter to describe it. It is termed the Kalb-Ramond (KR) (or ``gravitational'') axion field, to distinguish from the ordinary axion fields $a(x)$ that can play the r\^ole of dark matter in particle physics models.

We mention at this point that background geometries with (approximately) constant background $H_{ijk}$ torsion, where Latin indices denote spatial components of the four-dimensional space-time, may characterise the early universe. In such cases, as we shall discuss in section \ref{sec:lepto}, the H-torsion background constitutes extra source of CP violation, necessary for lepotogenesis, and through Baryon-minus-Lepton-number (B-L) conserving processes, Baryogenesis, and thus the observed matter-antimatter asymmetry in the Universe~\cite{sarkarlepto1,sarkarlepto}. Today of course any torsion background should be strongly suppressed, due to the lack of any experimental evidence for it. Scenarios as to how such cosmologies can evolve so as to guarantee the absence of any appreciable traces of torsion today can be found in \cite{sarkarlepto} and will be briefly reported upon in sections \ref{sec:lepto} and \ref{sec:brane}.

In the remainder of this section, we shall consider the effects of the quantum fluctuations of torsion, which survive the absence of any torsion background.
An important aspect of the coupling  of the torsion (or KR axion) quantum field $b(x)$ to
the  fermionic   matter  discussed   above  is  its   shift  symmetry,
characteristic of an axion field. Indeed, by shifting the field $b(x)$
by  a constant:  $b(x)  \to b(x)  +  c$, the  action (\ref{brr})  only
changes by  total derivative terms, such  as $c\, R^{\mu\nu\rho\sigma}
\widetilde{R}_{\mu\nu\rho\sigma}$               
and $c\, F^{\mu\nu}\widetilde{F}_{\mu\nu}$.  These terms are irrelevant for the
equations  of motion and  the induced  quantum dynamics,  provided the
fields fall off sufficiently fast  to zero at space-time infinity. 
The scenario for  the anomalous  Majorana mass generation  through torsion proposed in \cite{mptorsion}, and reviewed here, 
consists of augmenting the  effective action (\ref{brr}) by terms that
break such a shift symmetry. To illustrate this last point, we  first couple the KR axion $b(x)$ to
another pseudoscalar  axion field $a(x)$.   In string-inspired models,
such  pseudoscalar   axion~$a(x)$  may  be  provided   by  the  string
moduli~\cite{arvanitaki} and play the r\^ole of a dark matter candidate.    The  proposed   coupling  occurs
through  a mixing  in  the kinetic  terms  of the  two  fields. To  be
specific, we consider the action (henceforth we restrict ourselves to right-handed Majorana neutrino fermion fields)
\begin{eqnarray} 
  \label{bacoupl}
    \mathcal{S} \!&=&\! \int d^4 x \sqrt{-g} \, \Big[\frac{1}{2}
      (\partial_\mu b)^2 + \frac{b(x)}{192 \pi^2 f_b}
      {R}^{\mu\nu\rho\sigma} \widetilde{R}_{\mu\nu\rho\sigma} 
      + \frac{1}{2f_b^2} J^5_\mu {J^5}^\mu + \gamma
      (\partial_\mu b )\, (\partial^\mu a ) + \frac{1}{2}
      (\partial_\mu a)^2\nonumber\\ 
&&- y_a i a\, \Big( \overline{\psi}_R^{\ C} \psi_R - \overline{\psi}_R
\psi_R^{\ C}\Big) \Big]\;, \qquad 
\end{eqnarray}
where $\psi_R^{\ C} = (\psi_R)^C$ is the charge-conjugate right-handed
fermion $\psi_R$, $J_\mu^5  = \overline{\psi} \gamma_\mu \gamma_5 \psi$
is the  axial current of  the four-component Majorana fermion  $\psi =
\psi_R  +  (\psi_R)^C$,  and  $\gamma$  is  a  real  parameter  to  be
constrained later on.   Here, we have ignored gauge  fields, which are
not of interest to us,  and the possibility of a non-perturbative mass
$M_a$  for the  pseudoscalar  field~$a(x)$.  Moreover,  we remind  the
reader that the {\em repulsive} self-interaction fermion terms are due
to the existence of torsion in the Einstein-Cartan theory.  The Yukawa
coupling $y_a$ of  the axion moduli field $a$  to right-handed sterile
neutrino  matter $\psi_R$  may  be due  to  non perturbative  effects.
These terms \emph{break} the shift symmetry: $a \to a + c$.

It is convenient to diagonalize  the axion kinetic terms by redefining
the KR axion field as follows:
$b(x) \rightarrow {b^\prime}(x) \equiv b(x) + \gamma a(x)$. 
This implies that the effective action (\ref{bacoupl}) becomes:
\begin{eqnarray} 
  \label{bacoupl2}
&& \mathcal{S} =  \int d^4 x \sqrt{-g} \, \Big[\frac{1}{2}
      (\partial_\mu b^\prime )^2 + \frac{1}{2} \Big(1- \gamma^2
\Big) \, (\partial_\mu a)^2
  \nonumber \\ && + \frac{1}{2f_b^2} J^5_\mu {J^5}^\mu  +
\frac{b^\prime (x) - \gamma a(x)}{192 \pi^2 f_b}  
{R}^{\mu\nu\rho\sigma} \widetilde{R}_{\mu\nu\rho\sigma}  - y_a i a\, \Big( \overline{\psi}_R^{\ C} \psi_R - \overline{\psi}_R
\psi_R^{\ C}\Big) \Big]\;.\qquad
\end{eqnarray}
Thus we  observe that the $b^\prime  $ field has decoupled  and can be
integrated out  in the  path integral, leaving  behind an  axion field
$a(x)$  coupled   both  to  matter   fermions  and  to   the  operator
$R^{\mu\nu\rho\sigma}   {\widetilde   R}_{\mu\nu\rho\sigma}$,  thereby
playing now the  r\^ole of the torsion field.   We observe though that
the approach is only valid for
$
|\gamma|\ <\ 1\; , $
otherwise the  axion field would  appear as a  ghost, \emph{i.e.}~with
the  wrong  sign  of  its  kinetic  terms,  which  would  indicate  an
instability  of  the  model.  This  is the  only  restriction  of  the
parameter $\gamma$. In this case we may redefine the axion field so as to appear with a
canonical normalised kinetic term, implying the effective action:
{\small
\begin{eqnarray} 
  \label{bacoupl3} 
\mathcal{S}_a \!\!&=&\!\! \int d^4 x
    \sqrt{-g} \, \Big[\frac{1}{2} (\partial_\mu a )^2 - \frac{\gamma
        a(x)}{192 \pi^2 f_b \sqrt{1 - \gamma^2}}
      {R}^{\mu\nu\rho\sigma} \widetilde{R}_{\mu\nu\rho\sigma} \nonumber\\ 
&&\hspace{-5mm} - \frac{iy_a}{\sqrt{1 - \gamma^2}} \, 
a\, \Big( \overline{\psi}_R^{\ C} \psi_R - \overline{\psi}_R
\psi_R^{\ C}\Big) + \frac{1}{2f_b^2} J^5_\mu {J^5}^\mu
      \Big]\; .
\end{eqnarray}
\hspace{-1.5mm}}
Evidently, the  action $\mathcal{S}_a$ in~(\ref{bacoupl3}) corresponds
to a  canonically normalised axion field $a(x)$,  coupled \emph{both }
to the curvature of space-time, \emph{\`a la} torsion, with a modified
coupling $\gamma/(192 \pi^2  f_b \sqrt{1-\gamma^2})$, and to fermionic
matter  with  chirality-changing  Yukawa-like  couplings of  the  form
$y_a/\sqrt{1 - \gamma^2}$.
%******************************************************************
%Figure 1:  Anomalous Majorana mass generation
%******************************************************************
\begin{figure}[t]
 \centering
  \includegraphics[clip,width=0.40\textwidth,height=0.15\textheight]{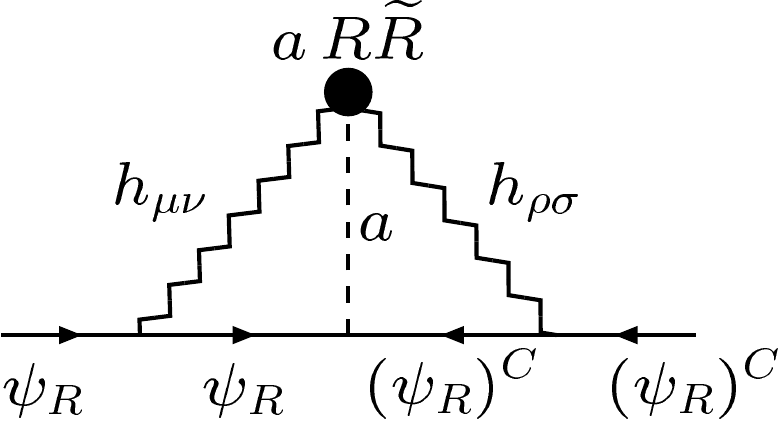} 
\caption{\it Typical Feynman graph giving rise to anomalous fermion
  mass generation~\cite{mptorsion}.  The black circle denotes the operator $a(x)\,
  R_{\mu\nu\lambda\rho}\widetilde{R}^{\mu\nu\lambda\rho}$ induced by
  torsion. The fields $h_{\mu\nu}$ denote graviton fluctuations.}\label{fig:feyn}
\end{figure}

The mechanism for  the anomalous Majorana mass generation  is shown in
Fig.~\ref{fig:feyn}.   We  may  now  estimate  the  two-loop  Majorana
neutrino mass in  quantum gravity with an effective  UV energy cut-off
$\Lambda$.    Adopting  the   effective   field-theory  framework   of
\cite{Donoghue:1994dn},  the gravitationally induced Majorana mass for right-handed neutrinos, $M_R$, is estimated to be:
\begin{equation}
  \label{MR}
M_R \sim \frac{1}{(16\pi^2)^2}\;
\frac{y_a\, \gamma\  \kappa^4 \Lambda^6}{192\pi^2 f_b (1 - \gamma^2 )} =
\frac{\sqrt{3}\, y_a\, \gamma\,  \kappa^5 \Lambda^6}{49152\sqrt{8}\,
\pi^4 (1 - \gamma^2 )}\; .  
\end{equation} 
In a UV
complete theory  such as  strings, the cutoff $\Lambda$  and the Planck mass scale $M_P$  are related.

It is interesting  to provide a numerical estimate  of the anomalously
generated Majorana mass $M_R$. Assuming that $\gamma \ll 1$, the size 
of $M_R$ may be estimated from~(\ref{MR}) to be
{\small
\begin{equation}
  \label{MRest}
M_R \sim (3.1\times 10^{11}~{\rm GeV})\bigg(\frac{y_a}{10^{-3}}\bigg)\;
\bigg(\frac{\gamma}{10^{-1}}\bigg)  
\bigg(\frac{\Lambda}{2.4 \times 10^{18}~{\rm GeV}}\bigg)^6\, .
\end{equation}
\hspace{-1.5mm}}
Obviously, the generation of $M_R$ is highly model dependent.  Taking,
for example, the quantum gravity  scale to be $\Lambda = 10^{17}$~GeV,
we  find that  $M_R$ is  at the  TeV scale,  for $y_a  =  10^{-3}$ and
$\gamma =  0.1$. However, if we  take the quantum gravity  scale to be
close  to the  GUT scale,  i.e.~$\Lambda =  10^{16}$~GeV, we  obtain a
right-handed neutrino  mass $M_R \sim  16$~keV, for the choice  $y_a =
\gamma = 10^{-3}$.   This is in the preferred  ballpark region for the
sterile   neutrino    $\psi_R$   to    qualify   as   a    warm   dark
matter~\cite{nuMSM}.

In a  string-theoretic framework, many  axions might exist  that could
mix  with each  other~\cite{arvanitaki}.  Such  a  mixing can  give rise  to reduced  UV
sensitivity of  the two-loop  graph shown in  Fig.~\ref{fig:feyn}.  In such cases, the anomalously  generated Majorana mass
may be estimated to be: \\
$M_R \sim 
\frac{\sqrt{3}\, y_a\, \gamma\,  \kappa^5 \Lambda^{6-2n} 
(\delta M^2_a)^n}{49152\sqrt{8}\, 
\pi^4 (1 - \gamma^2 )}\;$
for $n \leq 3$, and 
$ M_R \sim 
\frac{\sqrt{3}\, y_a\, \gamma\,  \kappa^5 (\delta M^{2}_a)^3}{49152\sqrt{8}\, 
\pi^4 (1 - \gamma^2 )}\; \frac{(\delta
  M^{2}_a)^{n-3}}{(M^2_a)^{n-3}}\; $
for  $n >  3$.  

It  is then  not difficult  to see  that  three axions
$a_{1,2,3}$ with next-to-neighbour mixing  as discussed above would be
sufficient  to obtain a  UV finite  result for  $M_R$ at  the two-loop
level. Of  course, beyond the two  loops, $M_R$ will  depend on higher
powers of the energy  cut-off $\Lambda$, i.e.~$\Lambda^{n> 6}$, but if
$\kappa\Lambda \ll  1$, these higher-order effects are  expected to be
subdominant.

In the above $n$-axion-mixing  scenarios, we note that the anomalously
generated  Majorana mass  term  will only  depend  on the  mass-mixing
parameters $\delta M_a^2$ of the  axion fields and not on their masses
themselves, as long as $n \le 3$. 
As a final comment we mention that the values of the Yukawa couplings $y_a$ 
may be determined by some underlying discrete symmetry~\cite{vlachos}, which for instance allows two of the 
right-handed neutrinos to be almost degenerate in mass, as required for enhanced CP violation of relevance to leptogenesis~\cite{sarkarlepto},
or in general characterises the $\nu$MSM~\cite{nuMSM}. These are interesting issues that deserve further exploration.

We stress at this point  that, although above we kept our discussion general, and included right-handed neutrinos in the spectrum of our models, assuming that the Yukawa-oridinary-axion interactions with them are the dominant ones, nevertheless our mechanism
applies also to models with \emph{no right-handed} neutrinos. Indeed, all one has to do in such cases is to assume that the ordinary-axion-Yukawa interactions occur in the active neutrino sector directly. At any rate, as in the next section our interest lies on producing Leptogenesis using (heavy) right handed neutrinos, from now on we do assume the existence of such heavy particles. However, as we shall argue below, for producing sufficient Leptogenesis using a KR torsion background in the early Universe, we need only one generation of heavy right handed Majorana (RHM) neutrinos. It goes without saying that the extension of our scenario to three RHM neutrino species, \emph{e.g.} as required in seesaw models~\cite{seesaw,Schechter:1980gr}, 
is straightforward. 

\section{Kalb-Ramond-Torsion Backgrounds in the Early Universe and Leptogenesis \label{sec:lepto}}

In this section we consider first the interacrtion of massive fermion species $\psi_i$, $i=1, \dots N$, of mass $m^{(i)}$,  with non-trivial gravitational and Kalb-Ramond H-torsion backgrounds, which we assume characterise the early Universe~\cite{sarkarlepto1,sarkarlepto}. From the considerations in the previous section \ref{sec:torsion}, it becomes clear that the physical content of the effects of an H-torsion background on Leptogenesis is contained in the following fermionic action (we start with Dirac fermions, the generalisation to Majorana ones, of relevance to us here, is straightforward and will be considered later on):
\bea
\label{diracb}
&& S_{Dirac}=\frac{1}{2} \int  \mbox{d}^4x\;\sqrt{-g}\, i\left(\overline{\psi}_i \, \gamma^{a}\overline \nabla_{a}\psi_i -(\overline \nabla_{a}\overline{\psi}_i) \, \gamma^{a}\psi_i 
+2im^{(i)} \overline{\psi}_i\psi_i\right) = \nonumber \\ && 
\int  \mbox{d}^4x\;\sqrt{-g}\, \overline{\psi}_i \, \left(i\gamma^{a}\partial_{a}+\widehat B_{d}\gamma^5\gamma^d-m^{(i)} \right)\psi_i  \equiv
 S_{Dirac}^{\rm free} - \int d^4x \sqrt{-g}\, \widehat B_\mu J^{5 \, \mu}~, 
\eea
where summation over the species index $i$ is implied, $J^{5 \, \mu} \equiv \overline \psi_i \, \gamma^\mu \, \gamma^5 \, \psi_i $, 
and the axial vector $\widehat B^{d}$ is defined by
\be\label{bddef}
\widehat B^{d}=\frac{1}{4}\varepsilon^{abcd}e_{a}^{\;\mu} \overline \omega_{b\mu c} = \frac{1}{4}\varepsilon^{abcd}e_{a}^{\;\mu}\, e_{b\nu}\left(\partial_{\mu}e_c^{\, \nu} + e^{-2\phi}H^{\nu}_{\;\mu\lambda} \, e_c^{\, \lambda}\right).
\ee
In this last step we have used  (\ref{omegabar}), (\ref{connection}) and the symmetry $\Gamma^\lambda_{\mu\nu} = \Gamma^\lambda_{\nu\mu}$ of the \emph{torsion-free} Christoffel symbol. In the special cases of interest here, either flat (Minkowski) or  Robertson-Walker space-times (which do not contain off-diagonal metric elements mixing temporal and spatial components), the axial vector $\widehat B^d$ is non-trivial and constitutes just the dual of the torsion tensor 
\be\label{defB}
\widehat B^d = -\frac{1}{4} \, \varepsilon^{abcd}\, e^{-2\phi}\, H_{abc}~.
\ee
In four space-time dimensions 
$\widehat B^\mu = \partial^\mu b~,$
where $b(x)$ is the Kalb-Ramond (KR) axion field (\ref{bfield}).

For a bosonic  string theory (with four uncompactified dimensions)  in non-trivial cosmological backgrounds, a world-sheet description has been  provided by a sigma model that can be identified with a Wess-Zumino-Witten type conformal field theory~\cite{Ellis-Nanopoulos-etc}.This construction has led to exact solutions (valid to all orders in the Regge slope, $\alpha^\prime$) for cosmological  bosonic backgrounds  with non-trivial metric, antisymmetric tensor and dilaton fields. Such solutions, in the Einstein frame, consist of (i) a Robertson-Walker metric with a scale factor  $a(t) \sim t$  where $t$ is the cosmic time, (ii) a dilaton field  $\phi$ that scales as   $\phi (t) \sim -{\rm ln}a(t), $ and (ii) a KR axion field scaling linearly with the cosmic time, $\overline {b} \propto t$ with $\overline b$ denoting the background value of $b$ (\emph{cf}. (\ref{bfield}). The resulting background axial vector $\overline{{\widehat B}}^d$ has only a non-trivial temporal component         
\begin{equation}\label{cosmolsol}
\overline b \sim {\rm const}\,  t ~, \qquad \partial^\mu \overline b \sim \epsilon^{\mu\nu\rho\sigma}\, e^{-2\phi} \, H_{\nu\rho\sigma}~,
\end{equation}and
\begin{equation}\label{constbdot}
\overline{{\widehat B}}^0 \propto {\dot {\overline b}} = {\rm constant}  \ne 0, \qquad  \overline{{\widehat B}}^{\,\, i} = 0, \, \quad  i=1,2,3~, 
\end{equation}
in the Robertson-Walker frame. 

In the absence of fermions, such constant H-torsion background in a Robertson-Walker space time \emph{do not} constitute solutions  equations of motion 
derived from the perturbative 
(in $\alpha^\prime$) action (\ref{low energy effective action}) (\emph{cf}. Appendix). 
However, in the presence of fermions coupled to the torsion $H$-field as in (\ref{diracb}), the  four-dimensional low-energy effective action 
 gives the following equations of motion  for the graviton and antisymmetric tensor:
\begin{eqnarray}\label{eqsmot}
&& {\rm graviton:}  \qquad R_{\mu\nu} - \frac{1}{4} H_{\mu}^{\,\, \alpha\beta} H_{\nu\alpha\beta} = 8\pi G \, \Big(T^\psi_{\mu\nu} - \frac{1}{2} g_{\mu\nu} T^\psi + {\rm dilaton-derivative~ terms}  +\dots \Big)~,\nonumber \\
&& {\rm antisymmetric~tensor:}  \qquad  \partial^\mu \Big(\sqrt{-g} e^{-2\phi}\,  \big[ H_{\mu\nu\rho} - \epsilon_{\mu\nu\rho\sigma} J^{5\, \sigma} + \dots \,\big] \Big) = 0~,
\end{eqnarray}
where $\dots$ denotes higher order terms in $\alpha^\prime$ in the gravitational part of the action, $T^\psi_{\mu\nu} $ is the stress-energy tensor of fermionic matter and $T^\psi = g^{\mu\nu} T_{\mu\nu}^\psi$. 
There is of course an equation of motion for the dilaton which provides additional constraints for the background. In order to simplify the analysis we will assume a constant dilaton below. A full analysis is given in the Appendix.

It is conceivable, as argued in \cite{sarkarlepto}, that in the presence of high temperature and densities of fermions (relevant for the early universe),
a condensate of the temporal component of the fermion axial current may be formed, in which case 
one may have (Lorentz-violating) \emph{perturbative} fixed points corresponding to a constant $H$-torsion. Indeed, 
from the equation for the antisymmetric tensor field (assuming a constant dilaton),  (\ref{eqsmot}), we observe that it can be solved upon using the pseudoscalar dual field $\overline b$ defined in (\ref{cosmolsol}): 
\begin{equation}\label{jb}
\partial^\mu \Big(\sqrt{-g} \big[ \epsilon_{\mu\nu\rho\sigma} (\partial^\sigma {\overline b} -  \tilde c \,  J^{5\, \sigma}) + \mathcal{O}\Big((\partial \overline b)^3\Big) \,\big] \Big) = 0~,
\end{equation}
where $\tilde c$ is a constant of proportionality.  From the above equations (in truncated form) it is clear that the fermion condensate can be a source of torsion .
Hence the non-perturbative solution (\ref{constbdot}),  derived in \cite{Ellis-Nanopoulos-etc} is still qualitatively valid since, from (\ref{jb}), we have
\begin{equation}\label{constbdot2}
  \dot{\overline{b}} =  {\tilde c}\langle J^{5}_0 \rangle =\tilde c  \, \langle \psi^\dagger_i \gamma^5 \psi_i \rangle  =  {\rm constant} \ne 0 ~,
\end{equation}
where $i$ runs over appropriate fermion species.

In \cite{Perturbative} a calculation in strong coupling gauge theory supported the formation of axial vector fermion condensates. At weak gauge coupling the condensates cease to form. The gauge coupling and string coupling are related in  string model building  of the fundamental interactions~\cite{tseytlin}. If the dilaton, rather than being a constant becomes more negative with time (as in the explicit solutions from bosonic string theory considered in \cite{Ellis-Nanopoulos-etc}), the string coupling and gauge coupling decrease with time; so there will be a time (and a critical value of $g$, $g_{c}$) when the gauge coupling will be too weak to support a condensate. This is of course qualitative: currently we can only speculate that the value of $g_{c}$ is achieved in the era of leptogenesis.  For a fundamental mechanism we would need to be quantitative but this has not been achieved as yet. 
Nevertheless, we note that in late era of the Universe, when the 
axial current condensate becomes vanishingly small, from (\ref{jb}) we find 
(upon ignoring (as subleading) the higher order $\mathcal{O}\Big((\partial \overline b)^3\Big)$ terms), that 
the rate of change of the $\overline{b}$ field diminishes with the cosmic time as the cube of the scale factor
\begin{equation}\label{bscale}
\dot{\overline{b}} \sim 1/a^{3}(t)~. 
\end{equation}

At this stage we notice that there is another challenging problem  at the microscopic level related to the cosmological constant and the need for fine-tuning, a generally unsolved problem at present. The time dependent pseudoscalar, with constant rate (\ref{constbdot2}) induces a vacuum energy term of the type of a positive cosmological constant once fluctuations around the background are allowed (\emph{cf.} Appendix, Eq. (\ref{cht})). There are ways by means of which such positive contributions  can be cancelled by negative vacuum energy contributions; for instance, in brane world models propagating in higher-dimensional bulk space times, one may obtain negative contributions on the brane vacuum energy from the bulk dynamics~\cite{rizos,mavromatosDfoam}, as we shall discuss in section \ref{sec:brane}.  Such negative contributions do not constitute an instability in the presence of the H-torsion constant background as they are cancelled by it. 
The real issue, however, is how, once the fermion condensates have disappeared, the brane vacuum energy acquires the present tiny value consistent with observational cosmology. One possible way to achieve this is associated with the dynamics of time dependent dilatons, on which we make some speculative remarks in the last part of this section.
Hence, at present we do not have a microscopic derivation  but rather a microscopic motivation for postulating our H-torsion background. Of course such issues do not apply to the neutrino mass generation mechanism of section \ref{sec:torsion}, which assumes zero H-torsion backgrounds.

A final remark before moving onto a discussion of leptogenesis concerns the path integration of quantum fluctuations of the antisymmetric torsion background.
Such fluctuations need to be included for an accurate estimate of the energy budget of the universe. Indeed, as we have seen in section \ref{sec:torsion}, such a process leads to repulsive axial current four fermion itneractions in the action that may contribute to the energy budget, especially in eras where condensation of the fermion axial current  occurs. To take formal account of such flucgtuations, a split of the $\widehat B$ field can be made explicit into background $\overline{\widehat{B}}^\mu $ and quantum fluctuation ${\widehat {\cal B}}^\mu$ parts, 
$\widehat {B}^\mu = \overline{\widehat{B}}^\mu + \widehat{\cal{B}}^\mu$,
where the background satisfies (\ref{constbdot}). 
The result for the relevant factor of the path integral after integration over the quantum fluctuations $\widehat B$ reads  
\begin{equation}\label{binteg}
\mathcal{Z} \propto \  \int D\psi \, D\overline\psi  e^{i {\widetilde S}(\overline{\widehat{B}}) +  i\int d^4 x \, \sqrt{-g}\, \frac{3}{16}\, \kappa^2  J_\mu^5 \, J^{5 \, \mu}}\, 
\end{equation}
where $\widetilde S (\overline{\widehat{B}}) = S(\overline{\widehat{B}}) + S_{Dirac}^{\rm free} - \int d^4x \sqrt{-g}\, \overline{\widehat{B}}_\mu J^{5 \, \mu}$ is the action in the presence of the background torsion, given by the sum of (\ref{low energy effective action}), and (\ref{diracb}). For a more detailed discussion we refer the reader to ref.~\cite{sarkarlepto}.
The presence of a fermion axial condensate of the type (\ref{constbdot2}) implies in general a generalised background for the fermions in a Hartree-Fock approximation
\begin{equation}\label{bfd}
B_\mu \equiv  \Big( \overline{\widehat B}_{\mu} +
\frac{3\, \kappa^2}{8} \, \mathcal{F}_{\mu} \Big)~, 
\end{equation}
where, on account of our previous discussion (\emph{cf.} (\ref{constbdot}), (\ref{constbdot2})) only temporal components of $B_\mu$ will be non zero, with 
${\mathcal F}_0  = \langle \psi_\ell ^\dagger \gamma^5 \psi_\ell \rangle \equiv \rho_R - \rho_L \ne 0~, \, {\mathcal F}_i = 0$. The reader should notice that the sum over fermion species  $\ell = 1, \dots N$, \emph{does not} include Majorana neutrinos that yield zero contributions to the condensate~\cite{sarkarlepto}. Hence, it is the other fermion fields of the SM, such as quarks, that could contribute to such condensates. 

We commence our discussion on H-torsion-background induced leptogenesis 
by considering a minimal extension of the Standard Model, with one right-handed massive (of mass $M$) Majorana neutrino field in the presence of constant axial backgrounds, $B_\mu$ (\ref{bfd}). The right-handed neutrino sector of such a model is described by
\be
\label{Lagrangian interacting theory}
\mathcal{L}=i\overline{N}\slashed{\partial}N-\frac{m}{2}(\overline{N^{c}}N+\overline{N}N^{c})-\overline{N}\slashed{B}\gamma^{5}N-Y_{k}\overline{L}_{k}\tilde{\phi}N+h.c.
\ee
where $N$ is the Majorana field with mass $m$ (that can be itself generated dynamically by the fluctuations of the torsion field in the way  described in section \ref{sec:torsion}),  $L_{k}$ is a lepton field of the SM, with $k$ a generation index, and the adjoint of the Higgs field is defined by the relation
$\tilde{\phi}_i=\varepsilon_{ij}\phi_j $, with $i,j = 1,2$ are SU(2) group indices.
We note  that, since our primary motivation here is to identify the axial background field with the totally antisymmetric part of a torsion
background, one should also consider 
the coupling of the axial field $B_\mu$ to all other fermions of the SM sector, $\psi_j$ ($j$=leptons, quarks) via a \emph{universal} minimal prescription. Hence, the coupling with all fermionic species is the same : $\overline \psi_j\,\gamma^5 \, \slash{B}\, \psi_j $. 
Specifically, as we have seen above, the identification of the torsion background with a homogeneous and isotropic cosmological Kalb-Ramond field in a string-theory-inspired model will lead to axial backgrounds with non-trivial temporal components only 
\begin{equation}\label{temporalB}
B_0 = {\rm const} \ne 0~, \, B_i = 0 ~, i=1,2,3~.
\end{equation}
Since in SM the leptons have definite chirality , the Yukawa interactions can be rewritten as
\be
\mathcal{L}_{YUK}=-Y_{k}\overline{L}_{k}\tilde{\phi}\left(\frac{1+\gamma^{5}}{2}\right)N-Y_{k}^{*}\overline{N}\tilde{\phi}^{\dagger}\left(\frac{1-\gamma^{5}}{2}\right)L_{k}.
\ee
Using the properties of the charge conjugation matrix and the Majorana condition, it is again seen to be equivalent to
\be\label{last Lagrangian}
\mathcal{L}_{YUK}=-Y_{k}\overline{L}_{k}\tilde{\phi}\left(\frac{1+\gamma^{5}}{2}\right)N-Y_{k}^{*}\overline{L}_{k}^{c}\tilde{\phi}^{\dagger}\left(\frac{1-\gamma^{5}}{2}\right)N.
\ee
It should be noted  that the two hermitian conjugate terms in the Yukawa Lagrangian are also CPT conjugate. This is to be expected on the basis of the CPT theorem. In fact CPT violation is introduced only by interactions with the background field. In the absence of the background, the squared matrix elements obtained from tree level diagrams for the two decays would be the same~\cite{Kolb Wolfram}. From the form of the interaction Lagrangian in Eq. (\ref{last Lagrangian}), it is straightforward to obtain the Feynman rules for the diagrams giving the decay of the Majorana particle in the two distinct channels. It also allows us to use positive frequency spinors both for the incoming Majorana particle and for the outgoing leptons.

\begin{figure}
\includegraphics[width=0.4\columnwidth]{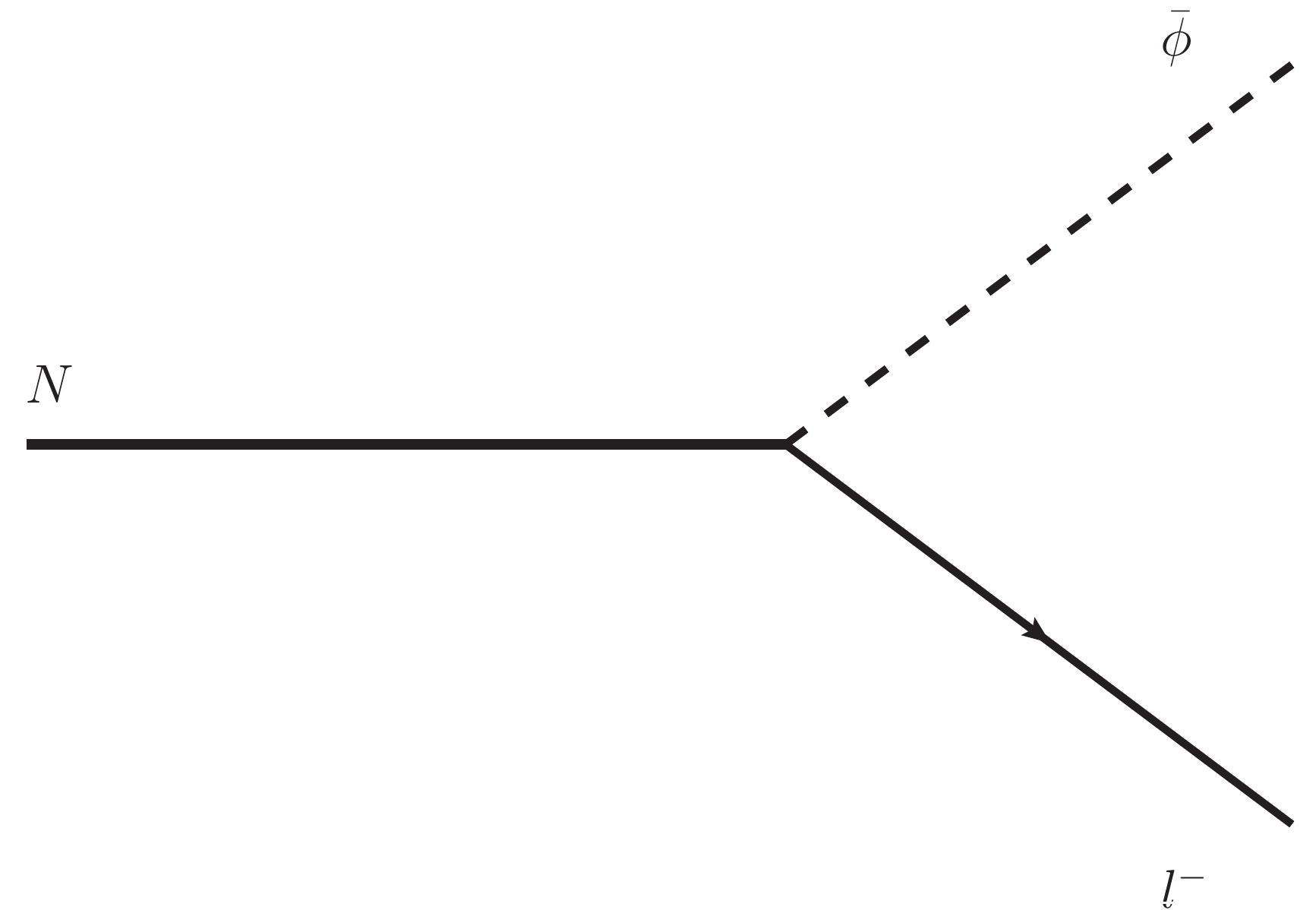} \hfill
\includegraphics[width=0.4\columnwidth]{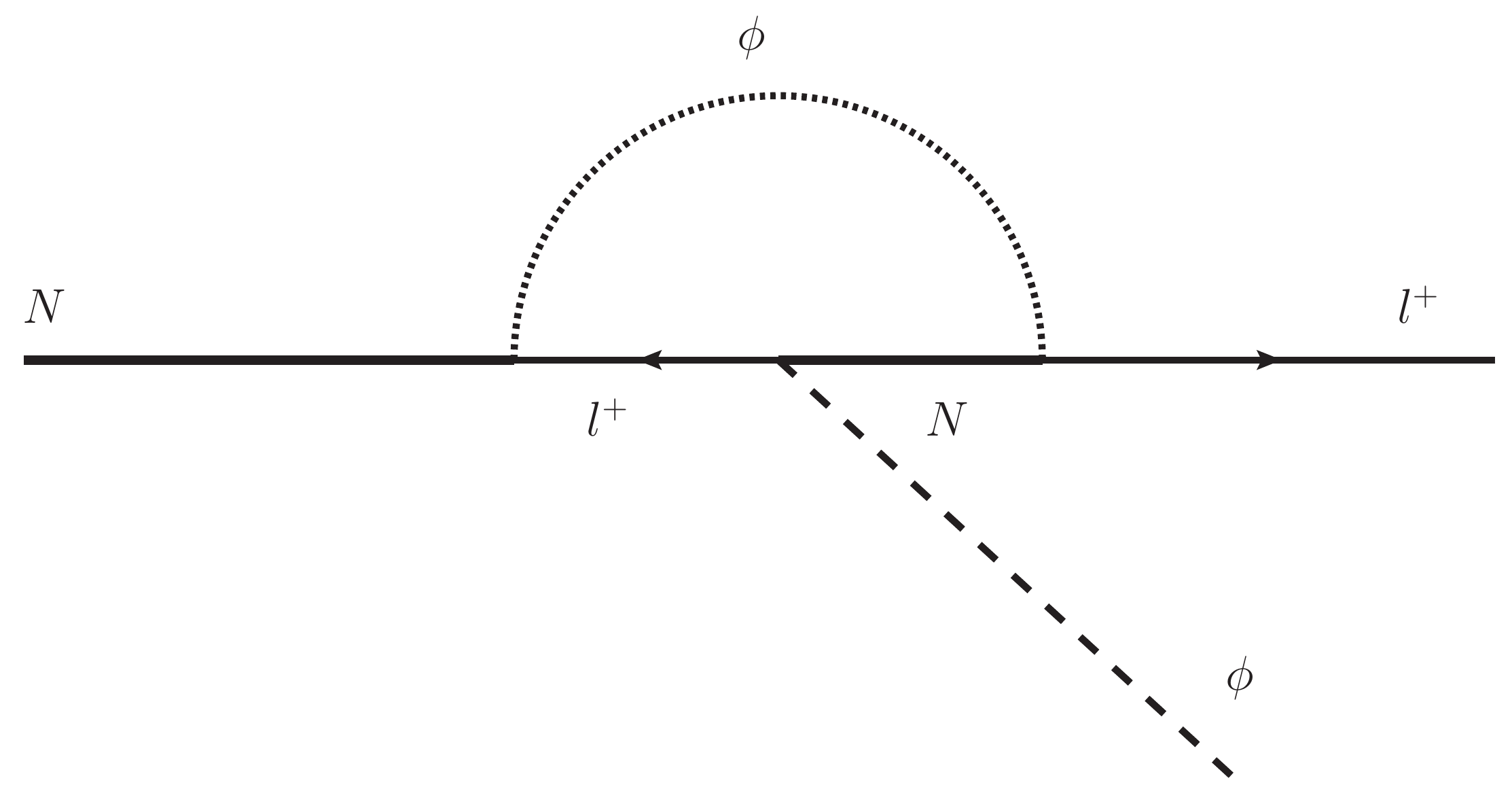} 
\caption{Tree- (left) and one-loop (right) decay amplitudes, corresponding to the Yukawa term that couples a right-handed neutrino to the standard model lepton sector. Analogous diagrams describe the decay in anti-leptons. Continuous undirected lines represent right-handed neutrinos, lines with an arrow are used to represent SM leptons,  whilst dashed lines correspond to the SM Higgs. The left diagrams are understood to be evaluated in the presence of a KR background field. The right diagram is the standard result of \cite{Fukugita-Yanagida}, leading to Leptogenesis.}
\label{fig:tree}
\end{figure} 
 Let us now turn to the study of the \emph{tree-level} decay processes of a Majorana right-handed neutrino into leptons and Higgs fields, depicted in (the left panel of) fig.~\ref{fig:tree}. For the qualitative purposes of the talk we assume that the massive right-handed neutrino is initially at rest. The more general case is studied in 
ref.~\cite{sarkarlepto}. For the channel $N\rightarrow l^{-}\overline{\phi}$ the decay rate is 
\be
\Gamma_1=\sum_k\frac{|Y_k|^2}{32\pi^2}\frac{m^2}{\Omega}\frac{\Omega+B_0}{\Omega-B_0}, \qquad \Omega=\sqrt{B_0^2+m^2}~,\ee
while, for the other channel, $N\rightarrow l^{+}\phi$,  the respective decay rate reads
\be
\Gamma_2=\sum_k\frac{|Y_k|^2}{32\pi^2}\frac{m^2}{\Omega}\frac{\Omega-B_0}{\Omega+B_0}.
\ee
The reader should notice that the decay rate of one process is obtained from the other upon flipping the sign of $B_0$.
The total decay rate is
\be
\Gamma=\Gamma_1+\Gamma_2=\sum_k\frac{|Y_k|^2}{16\pi^2}\frac{\Omega^2+B_0^2}{\Omega}.
\ee
It is worthwhile observing that this mechanism can produce a lepton asymmetry even with \emph{only one right-handed neutrino}, whereas the standard leptogenesis scenario \cite{Fukugita-Yanagida} requires at least three generations. Moreover, the occurrence of leptogenesis here is just due to decay processes at tree level, since the required $CP$ violation is introduced by the background field that enters in the external lines of Feynman diagrams.

The decay process goes out of equilibrium when the total decay rate drops below the expansion rate of the Universe, which is given by the Hubble constant \cite{Weinberg}
\be
\Gamma\simeq H=1,66 T^2 \mathcal{N}^{1/2} m_{P}^{-1}.
\ee
Here $\mathcal{N}$ is the effective number of degrees of freedom of all elementary particles and $m_{P}$ is the Planck mass.
From the last equation one can estimate the decoupling temperature $T_{D}$, in terms of the unknown parameters $\Omega$, $|Y|$ and $B_0$, is
\be
T_{D}\simeq6.2\cdot 10^{-2} \frac{|Y|}{\mathcal{N}^{1/4}}\sqrt{\frac{m_{P}(\Omega^2+B_0^2)}{\Omega}}.
\ee
In order for the inverse decay to be suppressed by the Boltzmann factor, we have to impose the further requirement that $T_{D}\leq \Omega$ when $\Gamma\simeq H$ (delayed decay mechanism \cite{Weinberg, Fukugita-Yanagida}). From this condition one can determine a lower bound for the mass $m$. In fact we are lead to the following inequality
\be
z(\Omega^2+B_0^2)\leq\Omega^3,
\ee
where $z=3.8\cdot 10^{-3}\frac{m_{P}|Y|^2}{\mathcal{N}^{1/2}}$. If we require that the bound is satisfied for all values of $B_0$ we get
\be\label{lowerb}
m^2\geq 1.09\, z^2.
\ee
 For us the Yukawa coupling  $Y$ is a free parameter.  If we assume $|Y|\approx 10^{-5}$, $\mathcal{N}\approx 10^2$, we get an order of magnitude estimate for the lower bound of the right-handed Majorana neutrino mass 
 \be
 m \ge  \overline{m}\approx 100 \;\mbox{TeV}~. 
 \ee
 Clearly in this mechanism for leptogenesis the low-mass right-handed neutrinos of \cite{nuMSM} do not fit. 

The lepton number density produced can then be estimated in the following way: by assumption all the neutrinos are at rest before the decay; hence the branching ratios of the decays are given by $r=\frac{\Gamma_1}{\Gamma}$ and $1-r$. The decay of a single neutrino produces the lepton number
\be \label{lepton number production in one decay}
\Delta L=r-(1-r)=2r-1=\frac{2\Omega B_0}{\Omega^2+B_0^2}.
\ee
Multiplying this quantity by the initial abundance of right-handed Majorana neutrinos at the temperature $T_D$ one gets an approximate estimate of the lepton number density. The density of the Majorana neutrinos is given by
\be \label{rhd}
n_{N}=\sum_{\lambda}\frac{1}{(2\pi)^3}\int\mbox{d}^3 p \, f(p,\lambda).
\ee
where as usual $\beta$ is the inverse temperature, $\lambda$ denotes the helicity and $f(p,\lambda)$ is the corresponding Fermi-Dirac distribution function. At high temperatures $T$, as the ones we are dealing with here,  this is well approximated by the Maxwell-Boltzmann function
$f(p,\lambda)=e^{-\beta\sqrt{m^2+(p+\lambda B_0)^2}}$. 
 It is now straightforward to see that, upon performing the sum over helicities in (\ref{rhd}), and the momentum integrals  for small $B_0 \ll m, T$~\cite{sarkarlepto},  one recovers the usual expression for the density of a non-relativistic species
$ n_{N}=e^{-\beta m}\,\left(\frac{m}{2\pi\beta}\right)^{\frac{3}{2}}+\mathcal{O}(B_0^2)$.

We assume that the right-handed neutrino density distribution follows closely the equilibrium distribution for $T\geq T_{D}$ and drops rapidly to zero at lower temperatures $T\leq T_{D}$; furthermore  the density of the sterile neutrino (normalised to the entropy density) is well approximated by a step-function. Therefore, upon multiplying (\ref{lepton number production in one decay}) by $n_{N}$,  we obtain that the total lepton asymmetry produced in the full decay of the right-handed neutrino is given by
\be\label{dltot}
\Delta L^{TOT}=(2r-1)n_{N}=\frac{2\Omega B_0}{\Omega^2+B_0^2}n_N
\ee
If the lepton asymmetry $\Delta L^{TOT}$ is communicated to the baryon sector by (Baryon (B))-(Lepton-number (L)) conserving sphaleron processes in the SM sector,  
then $\frac{\Delta L^{TOT}}{n_{\gamma}}$ is expected to be of the same order of magnitude of the (observed) baryon asymmetry for temperatures $T > 1$ GeV,
\be \label{baryon asymmetry}
Y_{\Delta B}=\frac{n_{B}-n_{\bar{B}}}{n_{\gamma}}=(6.1\pm0.3)\times10^{-10}
 \ee
where $n_{B}$ is the number density of baryons, $n_{\bar{B}}$ is the density of antibaryons in the universe, and $n_\gamma$ is the density of photons. An order of magnitude estimate of the ratio $\frac{B_0}{m}$ can then be found from (\ref{dltot}) by making use of the approximation $T_D\simeq m$ and retaining only first order terms in $\frac{B_0}{m}$. Recalling that the photon number density is
\be
n_\gamma \simeq \frac{2\zeta(3)}{\pi^2}\,T^3\simeq 0.24\, T^3
\ee
and that
\be
\frac{\Delta L}{n_\gamma}\simeq 10^{-10},
\ee
we estimate the ratio of the background field to the mass of the sterile neutrino to be
\be\label{largeB}
\frac{B_0}{m}\simeq 10^{-8}.
\ee
The small value of this ratio also allows us to justify \emph{a posteriori} the neglect of higher powers of $B_0$ in the formulae above. From the lower bound for the mass of 100 TeV found in (\ref{lowerb}), for the case where $Y = {\mathcal O}(10^{-5})$, we get an approximation for the smallest possible magnitude of the background field required in order for this mechanism to be effective $B_0\simeq1 \; \mbox{MeV}$.  
If other mechanisms contributed to the lepton asymmetry in the universe, or the Yukawa couplings assume smaller values, the minimum value of $B_0$ would be smaller than the one given here. Baryogenesis is then assumed to proceed via B-L conserving processes in the SM sector of the model.
In order to get a more accurate estimate of $B_0$,  the relevant Boltzmann equation will need to be studied. This requires a good approximation for the thermally averaged decay rates  of all the relevant processes and will be the subject of future research.

  We now remark that, in a microscopic model an $H$-torsion-induced background $B_0$ of the above magnitude of 1 MeV would correspond to a \emph{large positive} contribution to the cosmological constant in the effective action (\ref{binteg}) (\emph{cf.} Appendix, Eq.~(\ref{cht})); if uncancelled, this vacuum energy would modify the standard cosmology in the radiation dominated eras of the early Universe, on which the above estiamtes were based. There are possibilities whereby these fluctuations might be cancelled. As already mentioned, within the context of brane world quantum field theories  there may be anti-de Sitter type contributions from the bulk~\cite{rizos}. In such cases, there are negative vacuum energy contributions to the (four-dimensional)  brane vacuum. Such contributions may suppress the $B_0$-induced vacuum energy contributions to acceptable levels so that the standard cosmology may apply. 
 
 Moreover, in the context of axial condensates there is a phase transition whereby the gauge coupling is too weak for their formation \cite{Perturbative}. 
In the framework of our string cosmology, the string coupling $g_s = e^\phi$ 
is proportional to the gauge couplings, and is diminished with the cosmic time, due to the dilaton being a negative logarithmic function of the scale factor~\cite{Ellis-Nanopoulos-etc,lahanas}.  Thus it is likely  that, as the Universe cools down, the Lorentz-violating condensates, which require strong couplings and high densities in order to form, are destroyed via a suitable phase transition.  At this transition the $B$ field vanishes. In our scenario  this temperature needs to be lower than the decoupling one for leptogenesis. The destruction of the fermion condensate at a temperature $T \simeq T_D$, would imply that the Kalb-Ramond-torsion-axion field $b$ no longer varies linearly with the cosmic time but diminishes with the scale factor as in (\ref{bscale}). If one assumes a cooling law for the Universe of the form $a \sim T^{-1}$, then the $B^0$ torsion field would scale with the temperature  as $T^3$ for $T \le T_D$. Taking that into account,
the temperature of the Universe  today (from the CMB measurements) is $T_{\rm CMB} = 2.725~{\rm K} = 0.2348 ~{\rm m eV} $, and assuming that the perturbative fixed point is dominant at temperatures of the order of $T \simeq T_D = 100$~TeV, we obtain a cooling law for the torsion $B_0$-field of the form
\begin{equation}\label{b0today}
B_0 =  c_0 \, T^3~, \quad c_0 = {\rm 1~MeV}{(100~{\rm TeV})^{-3}} = 10^{-42}\, {\rm meV}^{-2}~.
\end{equation}
Thus, in such a scenario, the value of $B_0$  today would be of order 
\begin{equation}\label{Btoday}
B_{0 \, \rm today} = {\mathcal O}\Big(10^{-44}\Big)~{\rm meV}~, 
\end{equation}
which is much too small for any experimental detection, lying well within the upper bounds on $B_0 \le {\mathcal O}(10^{-2})$~eV,
placed by precision atomic experiments within the context of experimental tests of the Standard Model Extension~\cite{kostel,kostel2,bounds}.
These and other considerations require further evaluation. For the purposes of this work we will just consider the model with a background field which in the present era, away from the leptogenesis era, is effectively absent. Hence detailed models will need to confirm that the corresponding temperature, at which such a destruction can happen, is much higher than the O(100) GeV temperature at which the lepton asymmetry is transferred to baryon asymmetries due the B-L conserving processes in the SM sector of the model. This is an important issue for future study.

Another important issue is the effect of the constant antisymmetric torsion on the Cosmic Microwave  Background (CMB) radiation spectrum. If we assume that the CMB  spectrum is largely due to fluctuations at the surface of last scattering, which occurs at redshifts $z = \mathcal{O}(10^3)$, then we observe that at the corresponding temperatures $T = T_{\rm CMB} \, (1 + z) \sim 10^{3} \, T_{\rm CMB} $, the value of 
\begin{equation}\label{lastscat}
B_{0\, {\rm last~scat}} \sim 10^9 \, B_{0 \, \rm today} = {\mathcal O}\Big(10^{-35}\Big)~{\rm meV}~, 
\end{equation}
as follows from (\ref{b0today}). This is very small to produce any observable effects in the CMB spectrum as can be seen from the following argument:
one may consider higher derivative terms in the effective action of photons propagating in a torsion background (as is the case in string-effective theories). One then encounters, among others, higher-covariant-derivative-with-torsion terms of the form appearing in the Lorentz-violating electrodynamics~\cite{kostel2}, whose effects on cosmic microwave background radiation have been classified.  Among those terms are terms of the form $T^{\alpha\lambda}_{\,\,\,\rho} F_{\alpha\nu}\, \partial_\lambda {\tilde F}^{\rho\nu}$ and $T^{\sigma\gamma}_{\,\,\,\delta} F_{\sigma \nu}\, \partial_\gamma {F}^{\delta\nu}$, where $T^{\mu\nu\rho}$ is the torsion field and $\tilde F_{\mu\nu}$ is the dual of the photon field strength $F_{\mu\nu}$.
Such terms may be constrained by the mixing of electric (E-)and magnetic (B-) type polarizations of the CMB due to induced birefringence.  
With the strength of the Kalb-Ramond torsion in (\ref{lastscat}), the possible effects are well within the corresponding bounds. 

In general, the association of the Kalb-Ramond torsion with a pseudoscalar axion-like field, implies constraints of the interactions of this field with electromagnetic fields through the anomaly equation (\ref{anom}). In our model, the coupling of this interaction turns out to be the gravitational coupling, and thus such effects are very small, compatible with the current phenomenology.

 The cosmological vacuum energy, which is  proportional to $\exp\left(2\phi\right)$, also decreases with $t$ to very small values in string models, where relevant operators in a (perturbative) world-sheet renormalisation-group sense, drive the theory towards fixed points at which the dilaton field $\phi \sim -{\rm ln}a(t)$, where $a(t)$ is the scale factor of the Universe, which asymptotically in time $a \to \infty$~\cite{lahanas}. However, there are also some non-perturbative scenarios (with large string coupling) associated with brane worlds, which induce such phase transitions, but in which the dilaton tends to $+\infty$. This is considered in the next section. 

\section{Speculative Applications to Brane Models with Strongly-Coupled Strings \label{sec:brane}}

\begin{figure}[htb]
\begin{center}
\includegraphics[width=0.3\columnwidth]{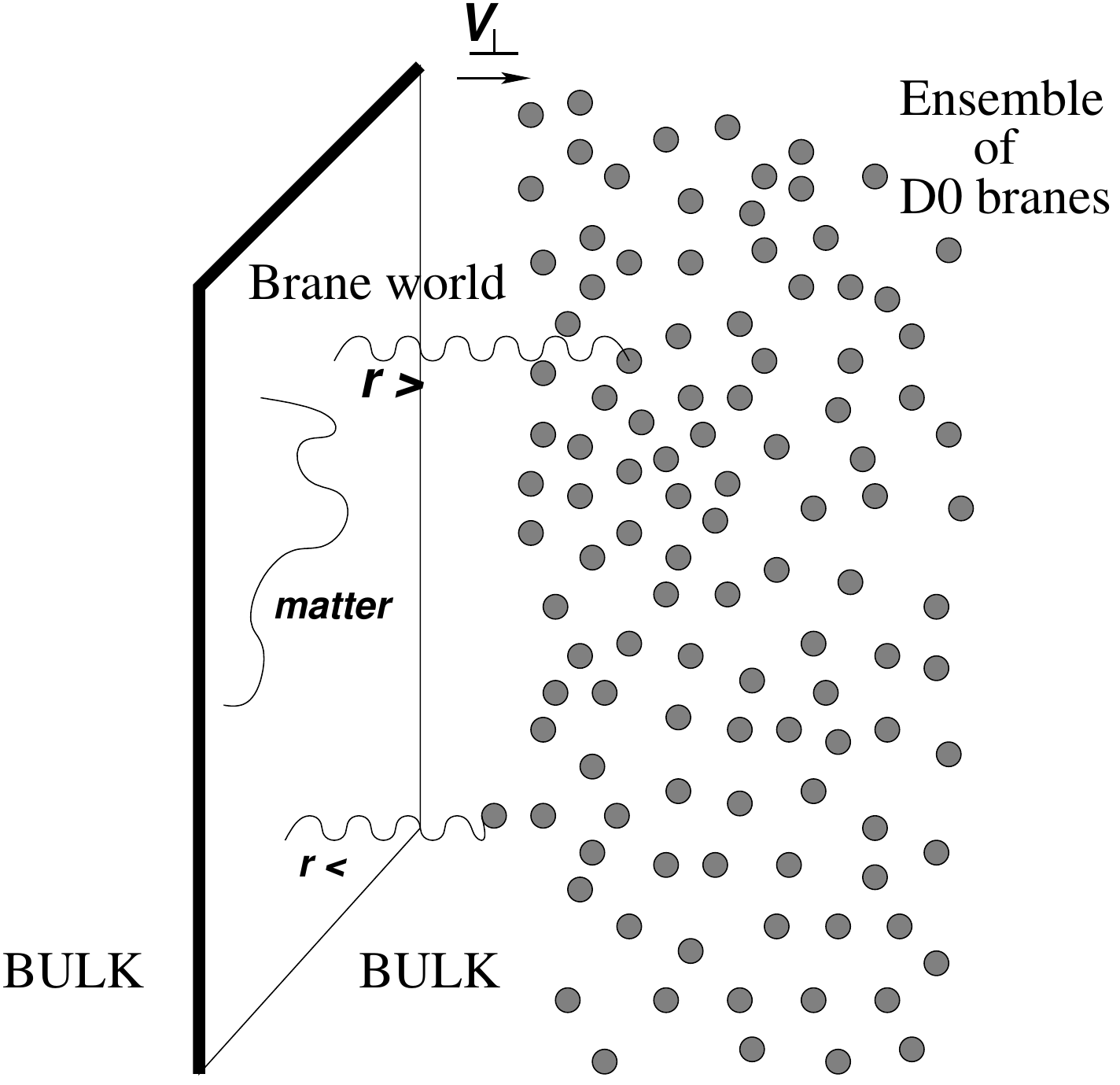}
\end{center}
\caption{A toy cosmological Brane-world model for the phase transition associated with the destruction of the Kalb Ramnod torsion background in the early Universe. A brane world moves adiabatically in the higher-dimensional bulk encountering at  a certain epoch, coinciding with the decoupling era of Leptogenesis, an ensemble of D0-brane defects of variable (in time) density, with respect to which the brane is assumed to move with velocity $V_\perp$. There are relative-velocity dependent interactions between the bulk defects and the brane world, described by open strings stretched between the brane and the defect. There are positive contributions to the brane's vacuum energy from defects at distances $r_{>}$ larger than the string length, and negative contributions from defects at distances $r_{<}$ shorter than the string length.  The latter  can lead to  bound states of D0-branes and the brane world which are then responsible for the destruction of the fermion condensates and the Kalb Ramond torsion.}
\label{fig:dfoam}
\end{figure}

We close the talk by presenting some speculations as to how the above scenarios of a rather large H-torsion background in early eras of the Universe, but negligible today,
can be realised in microscopic brane models of our Universe. Specifically, in this 
section, we argue that there can be solitonic degrees of freedom which can contribute to a Lorentz breaking background and be relevant in terms of the vacuum energy of the Universe at high energies. We consider a toy brane-world model.
The essential features of the model are depicted in fig.~\ref{fig:dfoam}. 
We consider a brane world (with three of the longitudinal brane directions uncompactified) moving in the bulk adiabatically, which encounters at a certain stage an ensemble of D0-brane defects (D-particles), which depending on the type of string theory can be either truly point-like or ``effectively'' point-like (e.g. in type II string theory one can have D3-branes wrapped up around three cycles, with a small radius of compactification, of the size of the string length).
Models of this type have been studied in \cite{mavromatosDfoam}, where we refer the reader for details. 

In what follows we shall only sketch the basic features from the point of view of interest in the current work, namely the phase transition at which the effects of the Kalb-Ramond torsion disappear. The brane universe representing our world moves in an inhomogeneous bulk space punctured on one side of the brane with 
an ensemble of D-particles. We assume that there is no cosmological expansion in the bulk, only along the longitudinal large directions of the brane universe one has such an expansion. Kalb-Ramond torsion constitutes an excitation restricted on the brane world, for the purposes of our toy example here. When the brane world is relatively far away from the ensemble of the defects then the brane vacuum energy is kept low due to cancelling effects between anti de Sitter contributions of vacuum energy in the bulk~\cite{rizos} and the constant torsion kinetic energy terms (\emph{cf.} Appendix (\ref{cht})).

As the brane world approaches the ensemble of defects, and is a few string lengths away~\footnote{There is a screening effect of the ensemble of D-particles due to the fact that open strings can be attached to the D-particles, which can thus cut macroscopic strings. 
 This  prevents any meaningful interactions between the D-particles in the ensemble and the brane world to extend 
beyond the order of a few string lengths.}, macroscopic open strings streteched between the brane and the D-particles are in operation, which imply gravitational potential energy contributions on the brane Universe, that are proportional to the 
sum of the squares of the components of relative velocity $v_\perp$ between the brane world and the D-particles transverse to the brane large longitudinal directions. 

As  discussed in \cite{mavromatosDfoam}, for bulk D-particles at a bulk distance $ r > \sqrt{\alpha^\prime}$, larger than the string scale $\sqrt{\alpha^\prime}$, one obtains positive constributions to the brane vacuum energy, while for bulk D-particles near the brane world, at distances $r < \sqrt{\alpha^\prime}$,
one has negative contributions to the brane's vacuum energy. In the example of $D0-D8$ brane interactions discussed in \cite{mavromatosDfoam}, for instance, one has for the long- and short-distance contributions to the potential energy of the brane world
\begin{eqnarray}\label{contr}
V_{D0-D8}^{r \, > \, \sqrt{\alpha^\prime}} = + \frac{r}{8\pi \, \alpha^\prime} {\vec v}_\perp^2 ~, 
\quad V_{D0-D8}^{r \, < \, \sqrt{\alpha^\prime}} = - \frac{\pi \alpha^\prime}{12 \, r^3 } {\vec v}_\perp^2 ~.
\end{eqnarray}
This example is only indicative to demonstrate the different sign of the various  contributions to the vacuum energy, for the case at hand of D3-brane universes one should appropriate compactify the pertinent brane worlds, but this will not affect qualitatively our arguments.
This difference in sign, then, implies that, one can arrange
for the densities of far away and nearby bulk D-particles,
which are not in general homogeneous, to be such that the
total contribution to the brane world vacuum energy is
always subcritical, so that issues such as overclosure of the
universe by a significant population of D-particle defects
can be avoided. This is rather generic, independent of the type of branes involved. 

There is a new minimal length in the problem, set by $r_{\rm min} \sim \sqrt{v_\perp} \sqrt{\alpha^\prime}$, which for adiabatic brane 
motion $|v_\perp| \ll 1 $ can be much smaller than the string length. 
The important point to notice is that the attractive short-distance contributions for $r < \sqrt{\alpha^\prime}$ can lead to bound states of D0 branes with the brane Universe.
The full non-perturbative string theory machinary has to be invoked in order to study in detail the formation and features of such bound states,
which we shall not discuss here. However, for our point of view, it suffices to mention that for strings in the neighborhood of such bound states the 
string coupling $e^{\langle \phi \rangle }$ 
is now modified to~\cite{kabat}
\begin{eqnarray}\label{gs}
{\rm exp}(2 \langle \phi \rangle ) = g_s^2 \Big( 1 + \frac{g_s}{(r/\sqrt{\alpha^\prime})^2} \Big)^{3/2}
\end{eqnarray}
where $g_s$ is the perturbative string coupling, far away ($r/\sqrt{\alpha^\prime} \to \infty$)  from the D-particle. From (\ref{gs}) it becomes obvious that for a string excitation near a D-particle bound to the brane, $r /\sqrt{\alpha^\prime} \to 0 $ the effective string coupling becomes pracically infinite.
From a $\sigma$-model point of view, the corresponding effective actions (\ref{low energy effective action}), (\ref{diracb}), 
describing the propagation of a matter open string (say a standard model fermion) on the brane world near a bound D-particle defect, 
would be characterised by a $\phi \to +\infty$ divergent dilaton field, and thus zero Kalb-Ramond torsion (\ref{bddef}). This implies that upon the encounter of the brane world with the ensemble of bulk D-pafrticles (fig. \ref{fig:dfoam}) a phase transition occurs, implying a trivial background for the Kalb-Ramond axion field, $b = 0$, on the four space-time dimensional world.
Moreover, the presence of D0-bound states would destroy the fermion condensate, since now the matter fields (represented by open strings)
would be attached to the defects, and as a result any long-range coherence would be destroyed.

After the destruction of the fermion axial condensate, the brane continues to move in a bulk region with an ensemble of D-particle defects (until the present era) 
and hence, the anti de Sitter contributions of the bulk vacuum energy that have been used to cancel the H-torsion contributions during the Leptogenesis era, can now themselves be cancelled by the 
long-range \emph{positive} contributions to the brane vacuum energy due to the bulk D-particles that are relatively 
far away from the brane world ($r > \sqrt{\alpha^\prime}$).   
In this way the vacuum energy of the brane world remains small during the late, post-Leptogenesis, eras of the Universe and the standard cosmology (e.g. $\Lambda$CDM model) applies. It goes without saying that, at this stage, we are far from being able to construct phenomenologically realistic cosmological models of the Universe in this framework, but it is encouraging that some toy models, such as the above-described one, exist.

\section*{Appendix: Perturbative world-sheet Fixed Points and constant  Kalb-Ramond Torsion in first quantised string models \label{sec:appendix}}

Here we shall  discuss the r\^ole of string matter in ensuring that  torsion (provided by the 
antisymmetric tensor field strength) remains constant in time, within the context of low-energy string effective field-theory actions. We shall consider perturbation theory in powers of $\alpha^\prime$ (the Regge slope). To lowest-order in $\alpha^\prime$, the  Einstein-frame effective action including matter fields ( in four-large-space-time dimensions) reads~\cite{tseytlin}
\begin{eqnarray}\label{leea}
S= \frac{1}{2\kappa^2}\int \mbox{d}^4x\;\sqrt{-g}\left(R-2\partial^{\mu}\phi\partial_{\mu}\phi-e^{-4\phi}H_{\lambda\mu\nu}H^{\lambda\mu\nu}-\frac{2}{3}\delta c\exp\left(2\phi\right)\right) + \int \mbox{d}^4x \, \sqrt{-g}\, e^{4\phi} {\mathcal L}_{\rm matter}~. \nonumber \\
\end{eqnarray}
where $\delta c$ denotes a central charge deficit~\cite{Ellis-Nanopoulos-etc}. 
It is important to realise that ${\mathcal L}_{\rm matter}$ contains the coupling of the antisymmetric tensor field strength $H_{\mu\nu\rho}$ with  the axial fermionic current, $J_\mu^5$; such terms do exist in our model as a consequence of the interpretation of the $e^{-2\phi} H_{\mu\nu\rho}$ in the Einstein term as contorsion (\emph{cf}. (\ref{bddef})). 
 ${\mathcal L}_{\rm matter}$, to lowest order in $\alpha^\prime$,  depends linearly on $-\frac{1}{4}\, \epsilon^{\mu\nu\rho\sigma}\, e^{-2\phi } H_{\mu\nu\rho}\, { \overline \psi} \gamma^\sigma\, \gamma^5 \psi $,  where $\psi$ are generic fermion fields.  
 Such terms have not been considered explicitly in the literature before.

The corresponding equations of motion for the graviton, antisymmetric tensor fields and dilatons read, respectively:
\begin{eqnarray}\label{eqsmot2}
&& R_{\mu\nu} - \frac{1}{2} g_{\mu\nu} R = \kappa^2 \Big(T_{\mu\nu}^{\rm matter} + T^H_{\mu\nu} + T^{\phi}_{\mu\nu} + T^c_{\mu\nu}\Big)~, \nonumber \\
&& \nabla_\mu \Big( e^{-4\phi} \, H^{\mu\nu\lambda} +  \kappa^2 \, e^{2\phi} \, \epsilon^{\mu\nu\lambda\sigma} \, J^5_\sigma \Big) = 0~, \nonumber \\
&& \Box \phi - \frac{1}{3} \, e^{2\phi} \, \delta c  +  e^{-4\phi} H_{\mu\nu\lambda} \, H^{\mu\nu\lambda}  + 2\kappa^2 \, e^{4\phi}{\mathcal L}_{\rm matter} ^{\ne H} - \frac{1}{4}\,  \kappa^2 \, e^{2\phi} \, \epsilon^{\mu\nu\lambda\sigma} \, H_{\mu\nu\lambda} \,  J_\sigma^5 = 0~, 
\end{eqnarray}
where $J_\mu = { \overline \psi} \gamma_\mu \, \gamma^5 \psi$ and ${\mathcal L}_{\rm matter}^{\ne H} $ denotes the part of the matter lagrangian independent of couplings to $H_{\mu\nu\rho}$.  $T^i_{\mu\nu}$ are the stress tensors pertaining to the various fields, $i={\rm matter}, H,\phi$ and $c$, with $T_{\mu\nu}^c = (\frac{1}{3} \, e^{2\phi}\, \delta c )\, g_{\mu\nu}$ the vacuum energy contribution.  
To these equations one should add the equations of motion for the matter fields, which are of no direct relevance for our cosmological considertations.

In four dimensions, in the absence of any matter fields, the equation for the antisymmetric tensor field is solved trivially by the replacement 
\begin{equation}\label{fibH}
H^{\mu\nu\lambda} = e^{4\phi} \epsilon^{\mu\nu\lambda\rho} \partial_\rho b~, 
\end{equation}
(where $b$ is the Kalb-Ramond pseudoscalar). It is then the (classical) Bianchi identity $\partial_{[\mu} H_{\mu\rho\lambda]} = 0$, where $[\dots ]$ indicate antisymmetrization of the indices that serves as a dynamical equation for the $b$ field~\cite{Ellis-Nanopoulos-etc}.   Indeed, contracting this identity with $\epsilon^{\mu\nu\rho\lambda}$ in the Einstein frame, and, on noting that 
$\epsilon_{\mu\nu\rho\sigma}\, \epsilon^{\mu\nu\rho\lambda} = 6 \, \delta^\lambda_\sigma $,  we obtain the following equation for $b$
\begin{equation}
\Box \, b + 4 \partial_\mu \phi \, \partial^\mu b = 0  \quad \Rightarrow \quad {\ddot b} + \Big(3 \frac{\dot a}{a} + 4{\dot \phi} \Big) \,  \, {\dot b}  = 0~.
\end{equation}
From this we observe that, to ensure a time-independent $\dot b$ solution, 
\begin{equation}
{\dot b }= K ={\rm constant}~,
\end{equation}
we must have a dilaton depending on the scale factor of the FRW Universe, $a(t)$, as
\begin{equation}\label{dilsol}
\phi = -\frac{3}{4} {\rm } {\rm ln}\, a ~.
\end{equation}
Substituting then in the dilaton equation (\ref{eqsmot2}) we obtain
\begin{equation}
-\frac{3}{2} \Big[ \frac{1}{2} \frac{\ddot a}{a} + \Big(\frac{\dot a}{a}\Big)^2 \Big] = \frac{1}{3} \frac{\delta c}{a^{3/2}} - 2 \frac{K^2 }{a^3} + \dots~.
\end{equation}
where the $\dots$ denote contributions from matter fields that do not contain $H$-torsion terms. 

This equation can be combined with the Einstein equations for the metric field
\begin{eqnarray}\Big(\frac{\dot a}{a}\Big)^2 = \frac{1}{3}\kappa^2 \sum_i \rho_i ~, \quad 
\frac{\ddot a}{a} + \frac{1}{2} \, \Big(\frac{\dot a}{a}\Big)^2 =  - \frac{1}{2} \kappa^2 \sum_i p_i ~,
\end{eqnarray}
where $\rho_i, p_i$ indicate the energy density and pressure of the various stress energy tensors, assumed to be  ideal fluids.

It is straightforward to see that there are no solutions compatible with a simple scaling of the scale factor, as in the radiation era -a conventional cosmology assumed in our leptogenesis scenario. 

However, in the presence of gauge fields, such as the electromagnetic potential, colour gauge fields \emph{etc}., it is known that the Bianchi identity in string theory is modified , such that in the Einstein frame
\begin{equation}\label{modified}
\epsilon^{\mu\nu\rho\sigma} \partial_\mu H_{\nu\rho\sigma} = \Big(\epsilon^{\mu\nu\rho\sigma}\, H_{\nu\rho\sigma}\,\Big)_{; \mu} = {\rm constant} \times F_{\mu\nu} F_{\rho\sigma} \, \epsilon^{\mu\nu\rho\sigma} \equiv {\rm constant} \times F_{\mu\nu} \, {\widetilde F}^{\mu\nu} ~.
\end{equation}
where semicolon denotes covariant derivative with respect to the torsion-free connection; (N.B. $\epsilon^{\mu\nu\rho\sigma}$, the tensorial density, is covariantly constant). The proportionality constant appearing on the right-hand-side has dimensions of inverse mass squared in our conventions (since it is proportional to $\alpha^\prime$); the precise numerical coefficients depend on the details of the underlying string/brane model. As mentioned in the text, Eq.~(\ref{modified}) is also related  to the axial anomaly (\ref{anom})~\cite{anomalies}. 

At this stage we would  like to make a comment on the consistency of the vacuum current condensates (\ref{constbdot2}) with the 
the anomaly equation (\ref{anom}) in a Robertson-Walker background, of interest to us. In such a case, by taking the vacuum expectation values on both sides of the equation, and assuming a Robertson-Walker space-time, we obtain
$ 3 H \langle J^{5\, 0} \rangle \propto \langle {\mathcal G}(\omega, A) \rangle~, \, \langle J^{5\, 0} \rangle = {\rm constant}~.$
This in turn implies that the presence of an axial current condensate would be associated with a condensate involving gauge and gravitational fields, which should scale with cosmic time as the Hubble parameter.  In string theory, whose effective action is characterised by highly non-linear terms in gauge field strength, such condensates are possible. On assuming the existence of a (dominant over gravitational) condensate of gauge fields $< F_{\mu\nu} \, {\widetilde F}^{\mu\nu} > $ \cite{Perturbative,Volovik} in the Leptogenesis era, it is easy to see 
from (\ref{fibH}) that 
the modified Bianchi identity reads
\begin{equation}\label{modbb}
{\ddot b} + 3 \frac{\dot a}{a} \, \dot b + 4 \dot \phi \, \dot b  \propto \, e^{-4\phi} < F_{\mu\nu} \, {\widetilde F}^{\mu\nu} > 
\end{equation}
which implies that a \emph{time independent solution} for $\dot b$ is compatible with a constant dilaton \footnote{Typically the behaviour of the dilaton can be parametrised \cite{lahanas} as $\phi={\Phi_{0}} \, {\rm ln}\left[a(t)\right]$ with $\Phi_{0}$ negative. $\phi$ is a slowly varying function of time which has to decay away before the epoch of Bing-Bang-nucleosynthesis, so as not to disturb its delicate conditions, and so in the era of leptogenesis (that precedes  nucleosynthesis in our model) we will for simplicity take $\phi$ to be a constant $\phi = \Phi_{0}$.}, as a perturbative fixed point;
To ensure a constant dilaton, and thus $\Box \, \phi = 0$ in the dilaton equation of motion (\ref{eqsmot2}), one needs to impose
\begin{equation}\label{constr}
\frac{1}{4} \, \kappa^2 \, e^{2\phi } \epsilon^{\mu\nu\lambda\rho} \, H_{\mu\nu\lambda}\, \langle J_\rho^5 \rangle = e^{-4\phi} H_{\mu\nu\rho}^2 - \frac{1}{3} \, e^{2\phi}\, \delta c  + \dots
\end{equation}
where $\dots$ include contributions from the rest of the matter lagrangian. For consistency with rotational invariance only the temporal component $\langle J_0^5  \rangle \ne 0$ of the vacuum current condensate is non-trivial.
During the leptogenesis era we assume a time-independent  condensate $\langle J_0^5  \rangle $. After that era, typically the condensate will vary with density \cite{Goda:2013bka} as a linear or higher power. Consequently, as the Universe expands, and the matter density decreases as $a(t)^{-3}$, the condensate $\langle J_\rho^5 \rangle$ will also fall-off.  Furthermore, as we have also noted earlier, the gauge coupling decreases with time due to the dilaton dependence of the string coupling and so becomes too weak to support a condensate~\cite{Perturbative}. Substituting (\ref{constr}) in the original action (\ref{leea}), we obtain a term of the form (the terms containing $\delta c \, e^{2\phi}$ cancel out):
\begin{equation}\label{cht}
S \ni \frac{1}{2\kappa^2} \int d^4 x \sqrt{-g} \{ - 3 \, e^{-4\phi_0} H_{\mu\nu\rho}^2 + \dots \}  = \frac{1}{2\kappa^2} \int d^4 x \sqrt{-g} \{ - 18 \, e^{4\phi_0} \, (\dot b)^2 + \dots \} 
\end{equation}
where $\phi_0$ is the constant dilaton value. The constant $\dot b$ is fixed to produce Leptogenesis, as discussed in section \ref{sec:lepto}, and thus such a term contributes \emph{positively} to the vacuum energy of the respective cosmic fluid. Such terms may be cancelled by bulk dynamics in, \emph{e.g.} brane world models~\cite{rizos,mavromatosDfoam}, 
where anti de-Sitter type contributions to the brane world vacuum energy independent of the dilaton may be induced from the bulk. One such scenario is discussed briefly in section \ref{sec:brane}. The issue is, nonetheless, wide open and much more work is needed before phenomenologically realistic models become available.

\end{document}